\documentclass[%
 reprint,
showpacs,
 amsmath,amssymb,
 prd,
]{revtex4-1}

\usepackage{graphicx}
\usepackage{dcolumn}
\usepackage{bm}
\usepackage{comment}
\usepackage{color,soul}
\usepackage{subcaption}

\def\HH{\mathcal{H}}
\def\NN{\mathcal{N}}
\def\SS{\mathcal{S}}
\def\zero{^{(0)}}
\def\pd{\partial}
\def\fii{\varphi}
\def\surfkappa{\kappa_{(\ell)}}
\def\dd{\mathrm{d}}
\def\eps{\varepsilon}
\DeclareMathOperator{\ii}{\mathrm{i}}

\def\Lie{\pounds}

\def\FF{\mathcal{F}}
\def\Fstar{{}^\star\! F}
\def\JJ{\mathcal{J}}

\begin{document}

\preprint{APS/123-QED}

\title{The Meissner Effect for weakly isolated horizons}

\author{Norman G\"urlebeck}
 \email{norman.guerlebeck@zarm.uni-bremen.de}
\affiliation{%
ZARM, University of Bremen, \\
Am Fallturm, 28359 Bremen, Germany\\
DLR Institute for Space Systems\\
Linzer Str. 1, 28359 Bremen, Germany
}%
\author{Martin Scholtz}
\email{scholtz@utf.mff.cuni.cz}
\affiliation{
 Institute of Theoretical Physics, Charles University, V Hole\v{s}ovi\v{c}k\'ach 2, 162~00 Prague, Czech Republic
}%

\date{\today}

\begin{abstract}
  Black holes are important astrophysical objects describing an end state of stellar evolution, which are observed frequently. There are theoretical predictions that Kerr black holes with high spins expel magnetic fields. However, Kerr black holes are pure vacuum solutions, which do not include accretion disks, and additionally previous investigations are mainly limited to weak magnetic fields. We prove for the first time in full general relativity that generic rapidly spinning black holes including those deformed by accretion disks still expel even strong magnetic fields. Analogously to a similar property of superconductors, this is called Meissner effect.

\end{abstract}

\maketitle


\section{\label{sec:intro}Introduction}


Black holes, as a final state of stellar evolution, are nowadays considered standard astronomical objects. Their existence is predicted by general relativity, supported both by strong theoretical arguments  \cite{Hawking-Penrose-1970} and observational evidence, most directly in the recent detection of gravitational waves \cite{Abbott-2016}. If they are not surrounded by matter, they are treated as Kerr--Newmann black holes characterized by their mass, spin and charge alone. This fundamental prediction of general relativity is known as the no-hair theorem for black holes \cite{Chrusciel-2012}, although three-hair theorem would be a better name. Additionally, the charge is usually neglected in astrophysical environments.

The black hole's spin is successfully measured using the continuum fitting and the iron line method \cite{Reis-2014,Risaliti-2013,McClintock-2013} possibly augmented by gravitational lensing. These methods require the presence of an accretion disk, which does not comply with the aforementioned assumptions of the no-hair theorem. The masses of the accretion disk, which are small compared to the black hole's mass, are typically assumed to yield negligible perturbations. Yet, if tests of the no-hair theorem are carried out, as suggested for future observatories like the Event Horizon Telescope \cite{Psaltis-2016,Johannsen-2016}, the admittedly small effects by the disk may become non-negligible.
To estimate such effects, it is prudent to treat black holes in a more general setting allowing for deviations from the Kerr geometry caused by additional matter in general relativity as it was recently started for the no-hair theorem in \cite{Guerlebeck-2015}. Naturally, this raises the question which other properties of Kerr black holes are universally holding for any black hole and which are sensitive to a possible accretion disk.

We will show for one important property -- the so-called Meissner effect -- that it is universal. The Meissner effect describes the property of black holes to expel any magnetic field if they become extremal, i.e.,\ if they have a maximal spin. This is especially interesting, since observations suggest that many supermassive black holes are almost extremal \cite{Reynolds-2013,McClintock-2013}. If the spin would exceed this threshold, the singularity inside the black hole would become naked and visible to distant observers, which is believed to be unphysical and, thus, prohibited as summarized in the cosmic censorship conjecture \cite{Penrose-1998}.


On the theoretical side, an analogy between black holes and thermodynamics emerged quite early in works of Bekenstein and Hawking \cite{Bekenstein-1973,Hawking-1976}. In particular, they found that the surface gravity $\kappa$ of a black hole plays the role of its temperature $T$ via $T = \kappa / 2\pi$, where we choose geometrical units in which $G = \hbar = c = 1$. The spin $a$ and the mass $M$ of a Kerr black hole in turn determine its surface gravity $\kappa=\sqrt{M^2 -a^2}/(2M(M+\sqrt{M^2-a^2}))$. For extremal Kerr black holes, where $a=M$, the surface gravity and, hence, the temperature vanish.

The analogy with thermodynamics can be carried further. In particular, extremal black holes, for which the temperature vanishes, expel external magnetic \emph{test} fields much like superconductors \cite{Wald-1974,Bicak-Dvorak-II,Bicak-Janis,King-1975}. In light of these similarities, the effect was dubbed \emph{Meissner effect}. It has been investigated for electromagnetic fields coupled to the gravitational field around Kerr--Newmann black holes \cite{Bicak-Dvorak-III}, for special exact models containing magnetic fields \cite{Karas-Vokrouhlicky-1991,Karas-Budinova-2000,Gibbons-2014,Bicak-Hejda,Bicak-Ledvinka}, and in string and Kaluza-Klein theory \cite{Gibbons-1998}. A relation between the Meissner effect and entanglement was also discussed \cite{Penna-2014-entanglement}.

From this theoretical treatment, one might be led to believe that the Meissner effect has consequences on the production efficiency $\eta$ of jets via the Blandford--Znajek process \cite{Blandford-Znajek-1977}: $\eta \propto a^2\,\Phi_{\mathrm{BH}}^2$, where $a$ is the spin of the black hole and $\Phi_{\mathrm{BH}}$ is the time-averaged magnetic flux \cite{Penna-2013,Tchekhovskoy2010,Tchekhovskoy-2011}. Faster rotating black holes are expected to produce jets more efficiently. However, a simultaneous decrease of the magnetic flux, as predicted by the Meissner effect, might counter-balance this behavior.
This conclusion rests, however, on the assumption that there is no matter in the vicinity of the horizon.
In contrast the authors in \cite{Komissarov-2007} did not assume vacuum electrodynamics but rather force-free electrodynamics as it is suitable for accreting black holes. In that case, the Meissner effect was found to have no effect on the jet creation, for other models allowing matter crossing the horizon, see \cite{Takamori-2011,Penna-2014}.
%
Based on the aforementioned results it is generally believed that the presence of matter suppresses the Meissner effect. Indeed, this conclusion is corroborated by the observation of the black hole GRS 1915+105, which has a spin of $a=0.98\pm 0.01$ and still creates jets efficiently \cite{Steiner-2012,Miller-2013}.

In contrast to these approaches, we assume vacuum in an arbitrarily small vicinity of the horizon to show analytically that the Meissner effect is a \emph{general} property of isolated black holes in general relativity.

\section{The Meissner effect for astrophysical black holes}

Since we wish to discuss the Meissner effect for more general black holes than those described by the Kerr metric, we use the quasi-local definition of a weakly isolated horizon (WIH). It describes horizons in equilibrium, i.e., currently no matter or radiation falls in \cite{Ashtekar-Krishnan-2004}. However, the WIH can be penetrated by electric and magnetic fields. The WIHs play also a role in loop quantum gravity \cite{Ashtekar-Baez-1998}. Note also that the thermodynamics of WIHs was developed \cite{Ashtekar-Lewandowski-2002}. In what follows, we assume a stationary and axially symmetric spacetime in a neighborhood of an uncharged WIH, which corresponds to a generic black hole in equilibrium. Moreover, we assume that sufficiently close to the WIH we have no matter, i.e., we have electrovacuum. Let us stress the facts that we neither assume the symmetries globally nor do we make any assumption whatsoever about the matter further out, in particular, about a possible accretion disk.

We assume that the space-time contains a WIH \cite{Ashtekar-Lewandowski-2002}, i.e.\ a non-expanding null hypersurface $\HH$ on which the Einstein-Maxwell equations are satisfied, equipped with the normal $\ell_a$; by definition, there is no flux of matter or radiation through the horizon. For the description of the space-time, we employ the Newman--Penrose (NP) formalism \cite{Newman-Penrose-1962} in which the main geometrical quantities are the spin coefficients, the Weyl scalars and the matter is described by the scalar projections of the energy-momentum tensor of an electromagnetic field. The null normal $\ell_a$ is necessarily tangent to the geodesics generating the horizon and satisfies $D \ell_b = \surfkappa\,\ell_b$, where $D = \ell^a \nabla_a$ and the constant $\surfkappa$ is the surface gravity of the WIH. $\surfkappa$ vanishes for extremal horizons.

We take the notation, the coordinate system and the metric of such an arbitrary black
hole as in \cite{Krishnan-2012}. Additionally, we use standard spherical coordinates $\theta$ and $\fii$ on the
topological 2-spheres foliating the horizon. In these coordinates, the intrinsic geometry of the 2-spheres is given by the metric conformal to a unit Euclidean sphere with the conformal factor $R=R(\theta,\fii)$,
\begin{align}\label{eq:2-metric}
  \dd s^2 &= R^2\left( \dd \theta^2 + \sin^2\theta \,\dd \fii^2\right).
\end{align}
At any point of such a 2-sphere there are exactly two null future-pointing directions: $\ell^a$ is tangent to the horizon and we denote the other one by $n^a$ and fix its scaling by $\ell_a n^a = 1$. We complete these vectors to a full NP null tetrad by introducing two complex null vectors $m^a$ and $\bar{m}^a$ satisfying $m_a \bar{m}^a = -1$ which span the tangent space of the sphere. The intrinsic connection compatible with the metric (\ref{eq:2-metric}) is encoded in the complex spin coefficient
\begin{align}
  a\zero &= \alpha\zero - \bar{\beta}\zero = m^a \bar{\delta} \bar{m}_a,
\end{align}
where $\delta = m^a \nabla_a$ on $\HH$ is given by
\begin{align}\label{eq:delta}
  \delta &= \frac{1}{\sqrt{2} R}\left( \pd_\theta + \frac{\ii}{\sin\theta} \pd_\fii \right),
\end{align}
The transformation
\begin{align}\label{eq:spin}
  m^a & \mapsto e^{\ii \chi} m^a
\end{align}
is called \emph{spin}, where $\chi$ is an arbitrary real parameter. It corresponds to a rotation in the tangent space of a 2-sphere. A quantity $\eta$ is said to have the \emph{spin weight} $s$ if it transforms as
\begin{align}
  \eta &\mapsto e^{\ii s\,\chi}\eta
\end{align}
under the spin (\ref{eq:spin}). For a spin $s$ quantity $\eta$ one defines the spin raising and lowering operators $\eth$ and $\bar{\eth}$ by
\begin{align}
  \eth \eta &= \delta\eta + s\,\bar{a}\zero\,\eta, &
                                                 \bar{\eth} \eta&= \bar{\delta}\eta - s\,a\zero\,\eta.
\end{align}
Following \cite{Krishnan-2012}, we extend vectors comprising the null tetrad off the horizon by conditions
\begin{align}\label{eq:parallel transport}
  \Delta n^a &= \Delta \ell^a = \Delta m^a = 0,
\end{align}
where $\Delta = n^a \nabla_a$. In terms of the spin coefficients, conditions \eqref{eq:parallel transport} imply
\begin{align}\label{eq:gamma nu tau}
  \gamma &= \nu = \tau = 0
\end{align}
everywhere in the neighborhood of the horizon.

The full space-time geometry on the horizon and in its neighborhood is a solution of a characteristic initial value problem with the initial data given on two intersecting null hypersurfaces. The first one is the horizon $\HH$, the other one is an arbitrarily chosen null hypersurface $\NN$ transversal to $\HH$ and intersecting $\HH$ in a 2-sphere $\SS_0$. The free data on the sphere $\SS_0$ consists of the
aforementioned function $R$, the values of the spin coefficients $\pi\zero$,
$a\zero = \alpha\zero-\bar{\beta}\zero$, $\lambda\zero$ and $\mu\zero$, the Weyl scalars $\Psi_2\zero$ and $\Psi_3\zero$ and the electromagnetic scalar
$\phi_1\zero$, where we use the notation of \cite{Newman-Penrose-1962,Krishnan-2012}. The real and imaginary part of $\phi_1\zero$ are the flux densities of the electric and magnetic field through the sphere $\SS_0$, respectively. The Weyl scalar $\Psi_2\zero$ determines the multipole moments of the
horizon \cite{Ashtekar2004}, namely, its real part is related to the mass and
its imaginary part is related to the angular momentum of the horizon. The functions $\Psi_2\zero, a\zero, \pi\zero$ and $\phi_1\zero$ are not independent but they are constrained by the equations
\begin{align}
  \Re \Psi_2\zero &= |a\zero|^2  - \frac{1}{2}( \delta a\zero +\bar{\delta} \bar{a}\zero )+ |\phi_1\zero|^2, \\
  \Im \Psi_2\zero &= - \Im \eth\pi\zero;
\end{align}
the spin weight of $\pi\zero$ is $-1$. Finally, the spin coefficients $\lambda\zero$ and $\mu\zero$ describe the extrinsic curvature of the horizon. In order to have a fully determined initial value problem, the Weyl scalar $\Psi_4$ and the electromagnetic scalar $\phi_2$ must be specified on the null hypersurface $\NN$.

Next, imposing the aforementioned symmetries, we require that, in the neighborhood of $\HH$, there exists a time-like Killing vector $K^a$ which equals  $\ell^a$ on the horizon. A necessary condition for the existence of such a Killing vector field is given by \cite{Stephani-2003}
\begin{align}
 \nabla_c \nabla_a K_b &= R_{abcd}\,K^d,
 \label{eq:necessary conditions}
\end{align}
In addition, we introduce the axial Killing vector $\eta^a$ which acquires the form $\eta^a = (\pd_\fii)^a$ \cite{Lewandowski-Pawlowski-2014}. The Killing equation \eqref{np:KE:deltaK2} (with $K$ replaced with $\eta$) then implies
\begin{align}
  a\zero &= - \frac{1}{\sqrt{2} \,R^2} \left(R' + R\cot\theta\right).
\end{align}

The electromagnetic field $F_{ab}$ is assumed to possess the same symmetries, i.e.\ we impose
\begin{align}\label{eq:Lie of Fab}
  \Lie_\eta F_{ab} &= 0,\quad \Lie_K F_{ab} = 0,
\end{align}
so that the electromagnetic NP quantities do not depend on the coordinates $v$ and $\fii$. In the electrovacuum case, the anti-self dual part of $F_{ab}$
\begin{align}
  \FF_{ab} &= \frac{1}{2}\left( F_{ab} + \ii \Fstar_{ab}\right)
\end{align}
is a closed form which, together with (\ref{eq:Lie of Fab}) implies that the 1-form
\begin{align}
  \JJ_a &= \FF_{ab} \eta^b
\end{align}
is also closed
\begin{align}\label{eq:dJ}
  \nabla_{[a} \JJ_{b]} &= 0.
\end{align}

Writing the conditions \eqref{eq:necessary conditions} and \eqref{eq:dJ} in the NP formalism and restricting them to the horizon we arrive at the constraints
\begin{subequations}\label{eq:constraints}
\begin{align}
 \surfkappa\,\lambda\zero &= \bar{\eth}{\pi}\zero  +\left( \pi\zero \right)^2 , \label{eq:pi}\\
\surfkappa\,\phi_2\zero &= \bar{\eth}\phi_1\zero + 2\,\pi\zero\,\phi_1\zero,  \label{eq:phi1}
\end{align}
\end{subequations}
where $\pi\zero$ and $\phi_1\zero$ have spin weights $-1$ and $0$, respectively. Additionally, Eq.\ \eqref{eq:pi} implies that the spin coefficient  $\lambda\zero$ is time-independent, as can also be inferred from substituting \eqref{eq:constraints} into the expression for $\lambda$ in \cite{Krishnan-2012}. The instructive but tedious calculations showing the validity of Eqs.\ \eqref{eq:constraints} for Kerr black holes will be presented elsewhere.

Subsequently, we prove the Meissner effect for uncharged black holes, i.e., we show that the magnetic flux across extremal, axially symmetric and stationary horizons vanishes. This is done by determining $\phi_1$ explicitly. For extremal horizons \footnote{A WIH is called extremal, if the normal $\ell_a$ is affinely parametrized, which is just a matter of the choice of $\ell_a$. Here, we fix $\ell_a$ so as to coincide with the Killing vector and, thus, the extremality has an invariant geometrical meaning.}, where we have $\surfkappa = 0$, Eq.\ \eqref{eq:pi} can be solved in terms of the free function $R$:
\begin{align}\label{eq:pi solution}
 \pi\zero(\theta) &= \frac{R(\theta)\,\sin\theta}{c_{\pi} + \sqrt{2}\int\limits_0^\theta R^2(\tilde{\theta})\,\sin\tilde{\theta}\,\dd \tilde\theta},
\end{align}
where $c_\pi$ is a complex integration constant. Now, the solution of Eq.\ \eqref{eq:phi1} reads
\begin{align}
 \label{eq:flux}
 \phi_1\zero = \frac{c_\phi}{\left( c_\pi + \sqrt{2} \int\limits_0^\theta R^2(\tilde{\theta})\,\sin\tilde{\theta}\,\dd\tilde{\theta}\right)^2},
\end{align}
where $c_\phi$ is another complex integration constant. The total electric charge $Q$ and magnetic charge $Q^\star$ of the black hole, which are not restricted at this stage, are then given by
\begin{align}\label{eq:charge}
Q + \ii Q^\star
&= \frac{2\,\sqrt{2}\,\pi\,c_\phi}{c_\pi+\sqrt{2}\int\limits_0^\pi R^2(\tilde{\theta})\,\sin\tilde{\theta}\,\dd\tilde{\theta} }.
\end{align}
Requiring that both charges vanish we get $c_\phi= 0$. This in turn means that the magnetic and electric flux density encoded in $\phi_1\zero $ vanish everywhere at the horizon, thereby, proving the Meissner effect.

It is worth mentioning that the symmetries were essential for our derivation. For a general WIH, $\phi_1\zero$ is part of the free, unconstrained data, showing that for the Meissner effect some symmetry is necessary.
Indeed, it was shown that specific non-axially symmetric magnetic test fields penetrate the horizon of an extremal Kerr black hole, see \cite{Bicak-Janis}. Thus, the Meissner effect does not hold in this case. On the other hand, as those authors point out, the test field they consider is, in fact, not the limit of a stationary electromagnetic field \cite{Pollock1977}. Hence, the stationarity might be the crucial of the two symmetries.

We also emphasize that the existence of the Killing vectors was assumed only in the neighborhood of the horizon, not in the entire space-time.

\section{The expulsion of the magnetic field from the horizon}

The proof given above shows that the magnetic flux across any part of the horizon vanishes for strictly extremal black holes. For the understanding of the Meissner effect, the transition from the non-extremal case to the extremal one is important. We depict it in Fig.\ \ref{fig:meissner} for a specific deformation of the Kerr black hole. We fix the deviation by choosing $\phi_2\zero$ as the spin-weighted spherical harmonic ${}_{-1}Y_{2,0}$ \cite{Goldberg-1967},
\begin{align}
 \phi_2\zero &= C\,{}_{-1}Y_{2,0},
\end{align}
with an arbitrary non-vanishing constant $C$, leaving the other quantities, including the mass, unchanged. For each value of $\surfkappa$, we solve Eq.\ \eqref{eq:phi1} numerically and calculate the electromagnetic field in the neighborhood of the black hole using the NP field equations \cite{Newman-Penrose-1962}. Finally, we plot the level sets of the magnetic and electric flux density, i.e., the imaginary and real part of $\phi_1\zero$ rescaled by $C^{-1}$, respectively. We choose the contours of the constant rescaled dimensionless magnetic flux density to be equidistant, with the difference between two neighboring contours being $10^{-2}$.

Fig.\ \ref{fig:meissner} and \ref{fig:electric meissner} clearly show that the lines of constant non-vanishing flux density are penetrating the horizon for $a/M < 1$ (under-extremal case) and are expelled in the transition $a/M\to 1$ (extremal case). In Fig.\ \ref{fig:field lines} we plot the magnetic field lines for different spins of the black hole.

For the visualization, we transform the spherical coordinates $r, \theta$ to the Cartesian ones by the usual relations $x = r\sin\theta, y = r\cos\theta$ in all figures.

\section{Jet creation efficiency}

\begin{figure*}
  \begin{subfigure}{0.29\textwidth}
    \includegraphics[width=\textwidth]{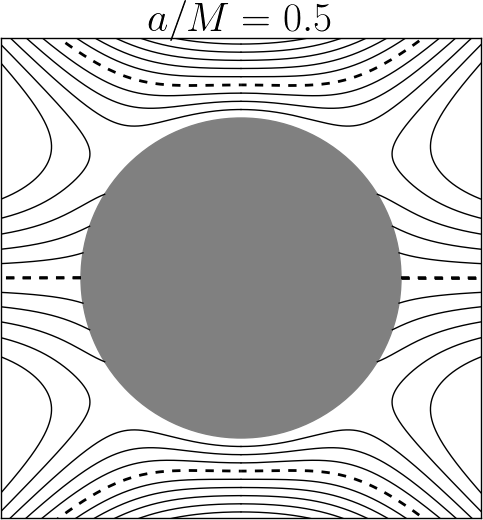}
  \end{subfigure}
  \begin{subfigure}{0.29\textwidth}
    \includegraphics[width=\textwidth]{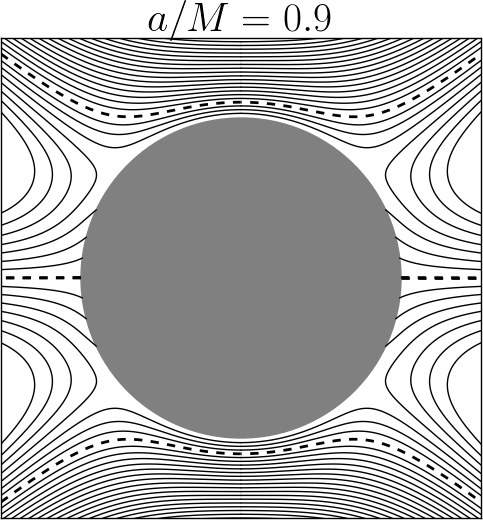}
  \end{subfigure}
    \begin{subfigure}{0.29\textwidth}
      \includegraphics[width=\textwidth]{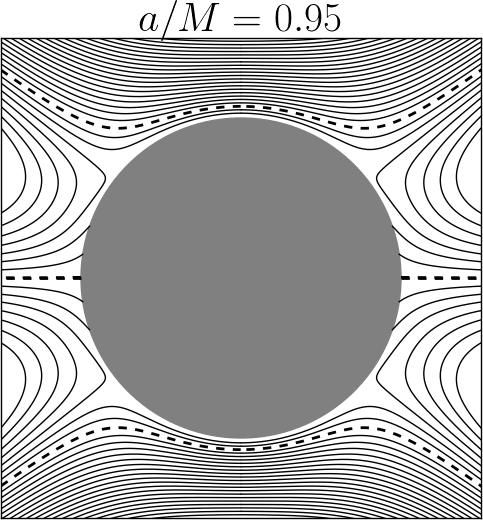}
    \end{subfigure}

    \begin{subfigure}{0.29\textwidth}
      \includegraphics[width=\textwidth]{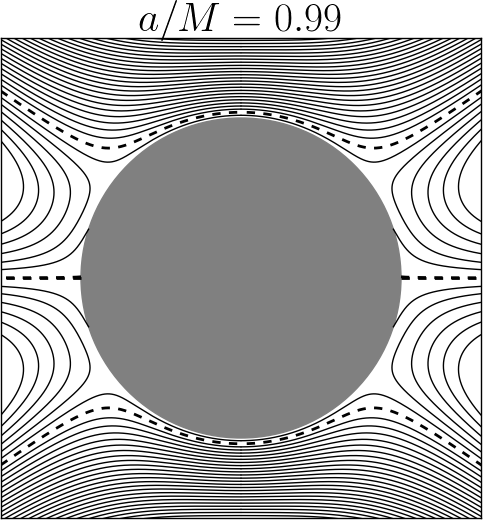}
    \end{subfigure}
    \begin{subfigure}{0.29\textwidth}
      \includegraphics[width=\textwidth]{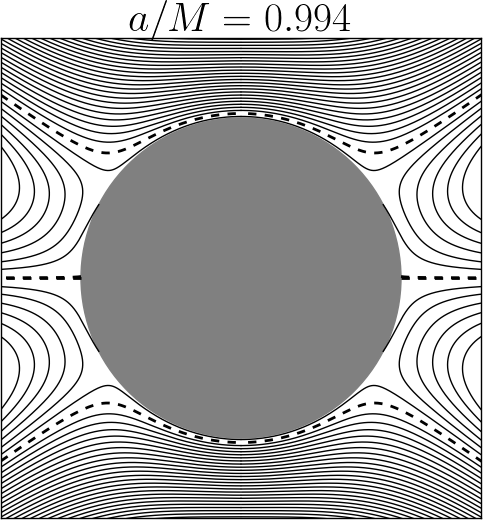}
    \end{subfigure}
    \begin{subfigure}{0.29\textwidth}
      \includegraphics[width=\textwidth]{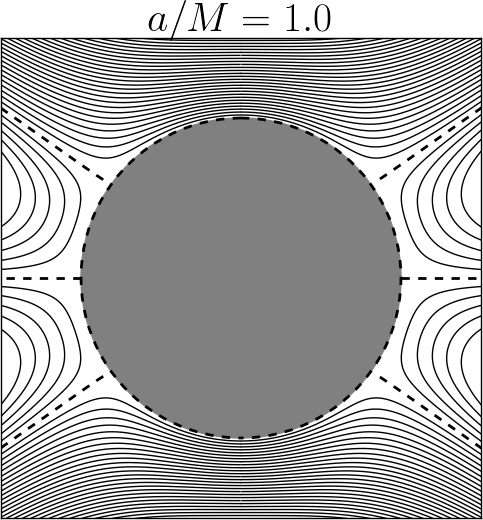}
    \end{subfigure}
\caption{{\bfseries Lines of equal magnetic flux density} for a given value of $a/M$. Dashed lines represent vanishing flux density.}
\label{fig:meissner}
\end{figure*}

\begin{figure*}
  \begin{subfigure}{0.29\textwidth}
    \includegraphics[width=\textwidth]{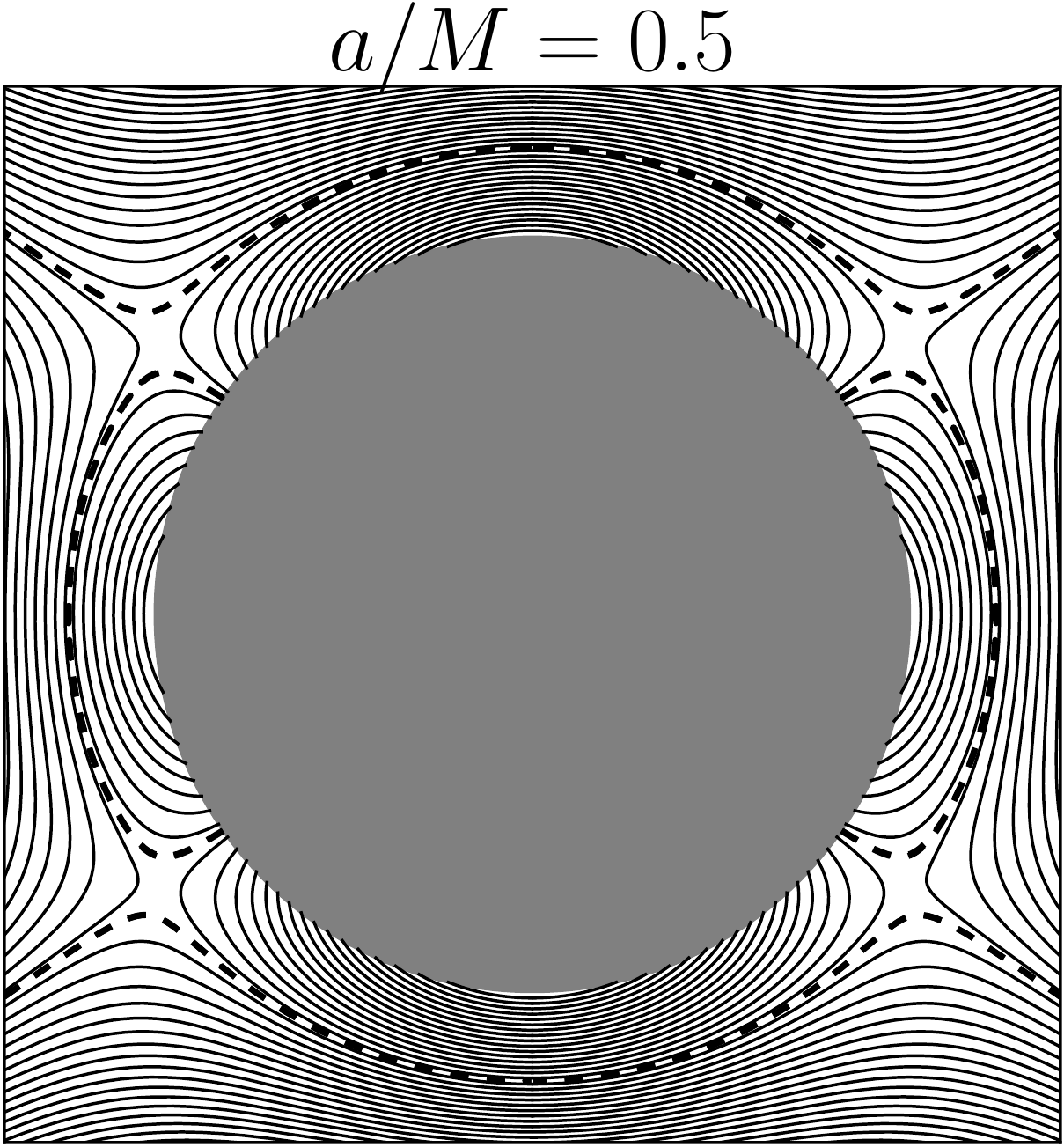}
  \end{subfigure}
  \begin{subfigure}{0.29\textwidth}
    \includegraphics[width=\textwidth]{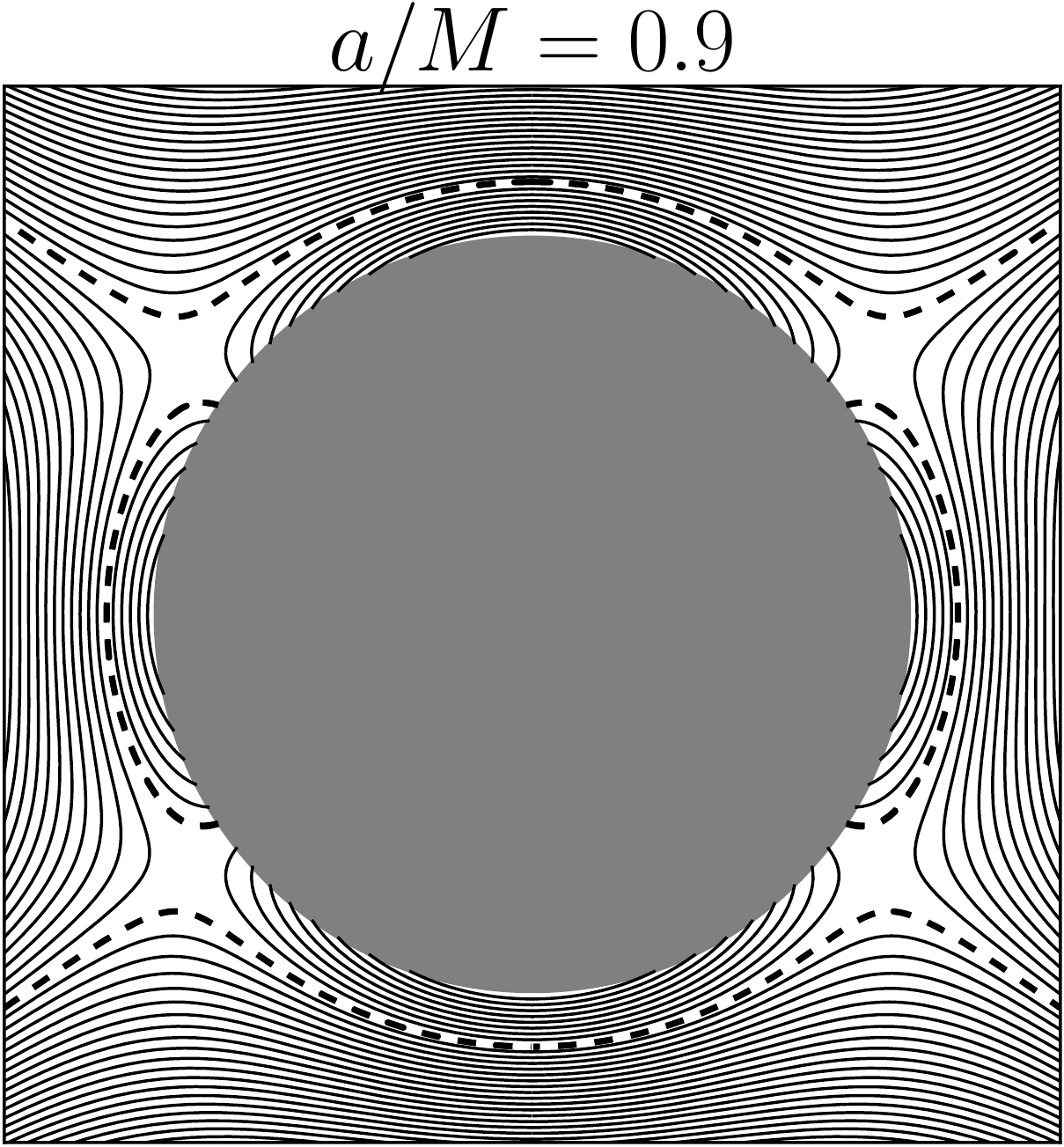}
  \end{subfigure}
  \begin{subfigure}{0.29\textwidth}
    \includegraphics[width=\textwidth]{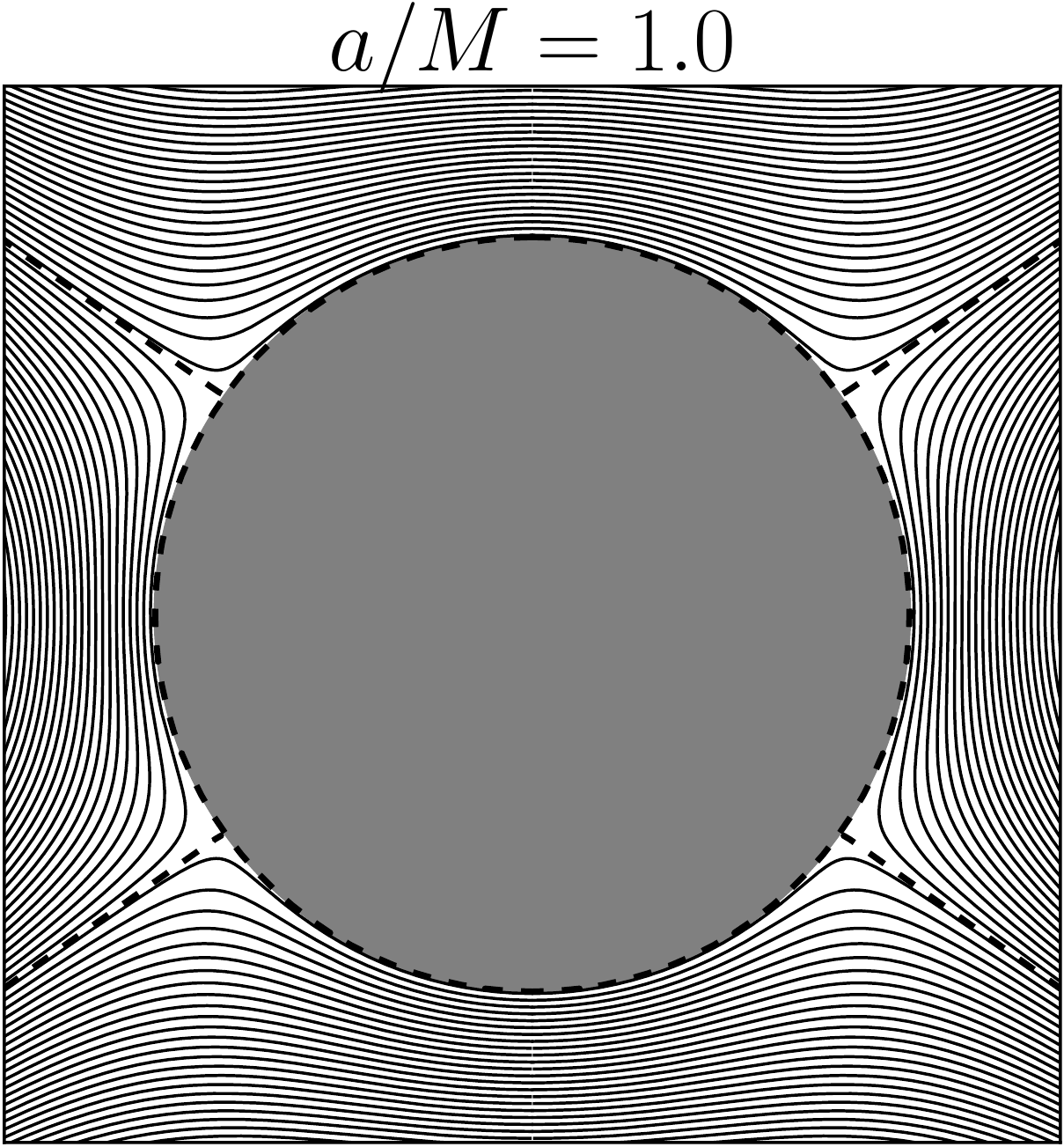}
  \end{subfigure}
\caption{{\bfseries Electric flux density}. The lines of equal electric flux
density around the black hole for a given value of the spin parameter $a/M$. Dashed
thick lines represent the lines of zero flux density. }
\label{fig:electric meissner}
\end{figure*}

\begin{figure*}
  \begin{subfigure}{0.29\textwidth}
    \includegraphics[width=\textwidth]{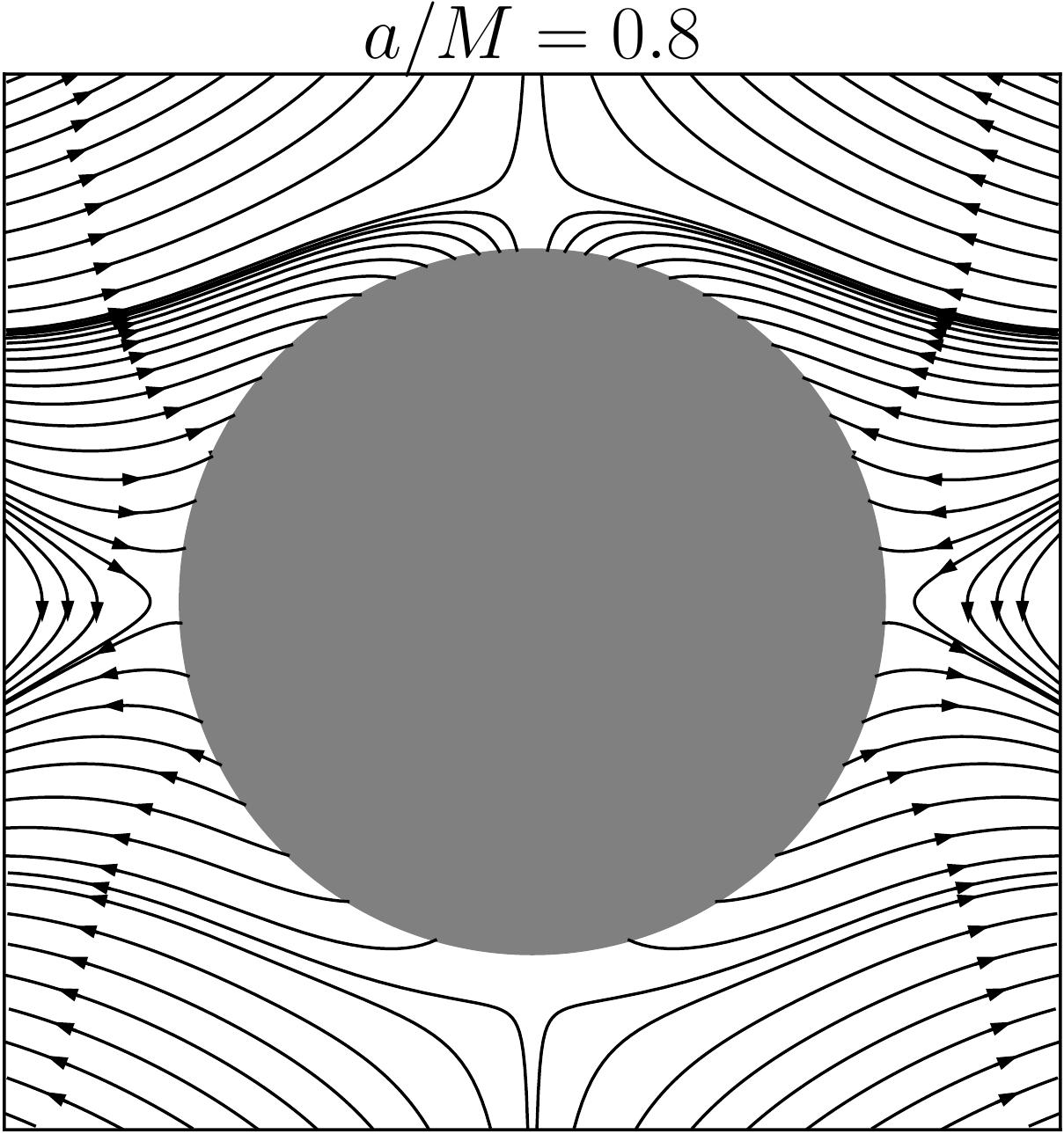}
  \end{subfigure}
  \begin{subfigure}{0.29\textwidth}
    \includegraphics[width=\textwidth]{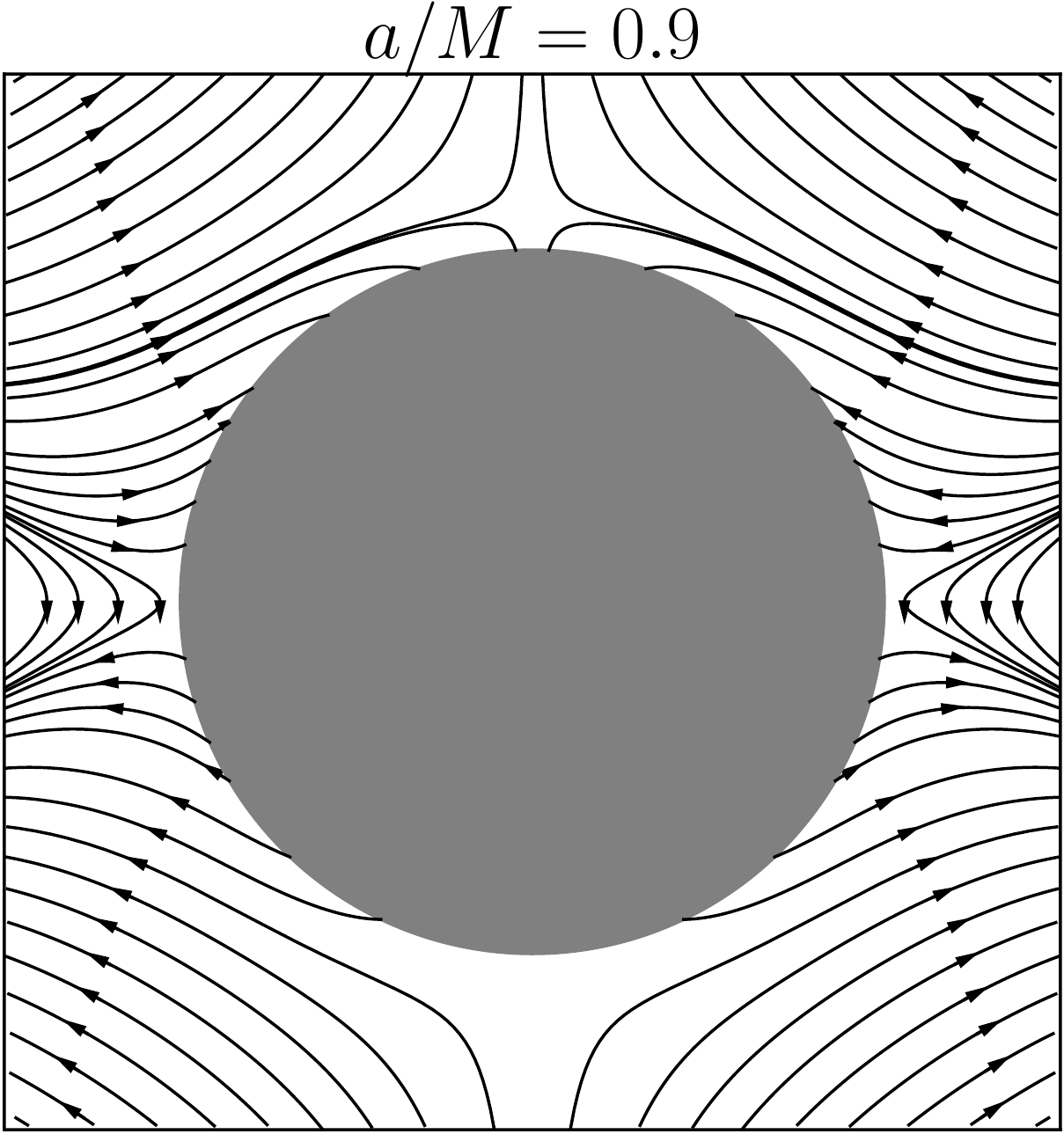}
  \end{subfigure}
  \begin{subfigure}{0.29\textwidth}
    \includegraphics[width=\textwidth]{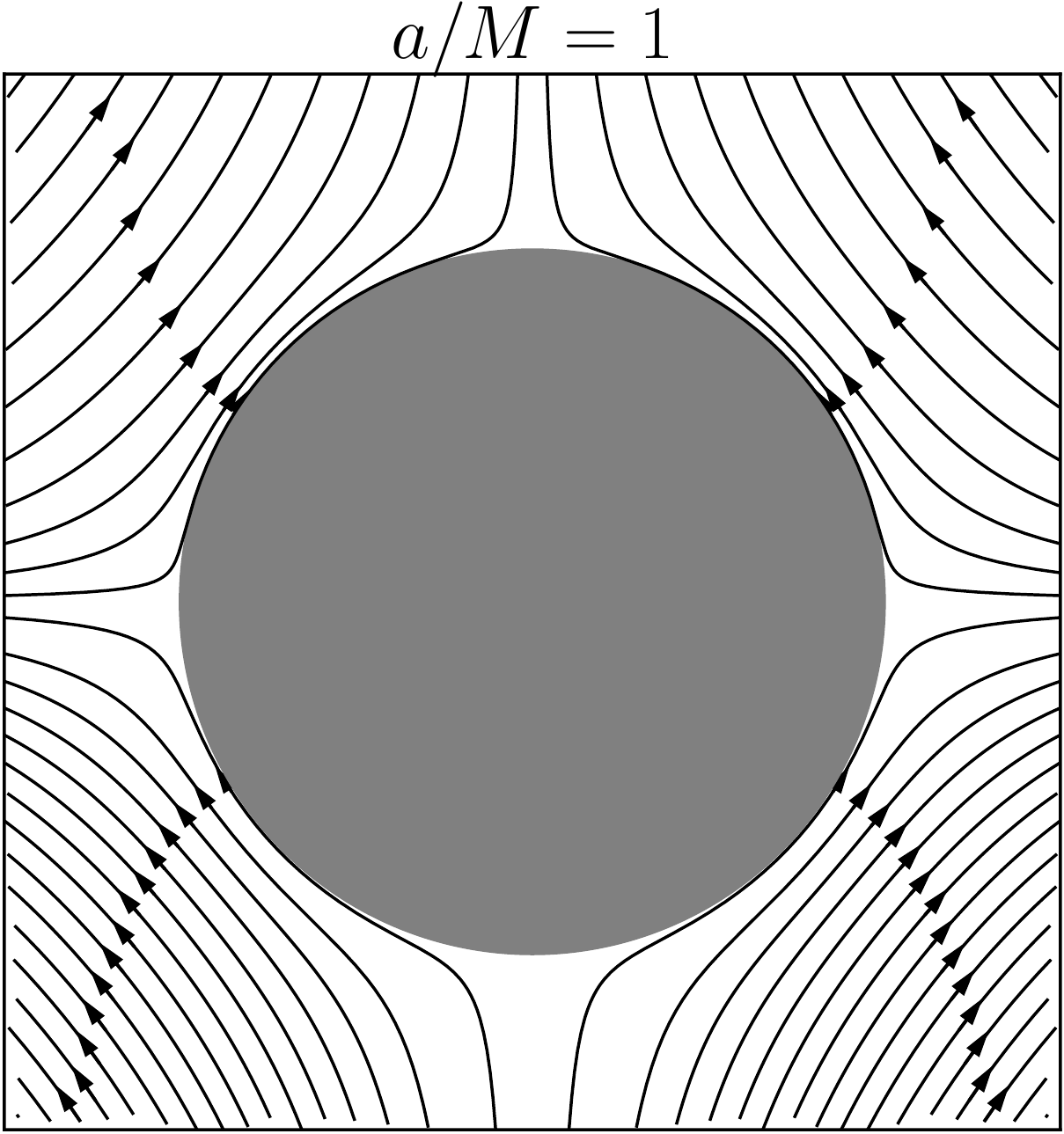}
  \end{subfigure}

  \caption{{\bfseries Field lines of the magnetic field} $B_a$ measured by an observer with the four-velocity $u^a = (\ell^a+n^a)/\sqrt{2}$, i.e., $B_a = {}^\star \! F_{ab} u^b$, where  ${}^\star\! F_{ab}$ is the dual of the electromagnetic field tensor defined by $\phi_0$, $\phi_1$ and $\phi_2$ \cite{Stewart1993}. }
 \label{fig:field lines}
\end{figure*}

Although our analysis and Figs.\ \ref{fig:meissner}--\ref{fig:field lines} show that the Meissner effect holds for generic black holes in equilibrium in general relativity, the impact on the jet creation
efficiency has still to be assessed. As we explained in the introduction, the Meissner effect does not operate in the presence of matter and the Blandford--Znajek process requires an influx of accreting matter through the black hole horizon, while we assumed the black hole to be isolated in our approach. Nevertheless, since the Meissner effect plays a role only in the limit of maximal spin, it will be interesting to see how strongly it could affect the jet creation efficiency. In order to do so, we assume here that the accretion influx, while powering the jet via the Blandford-Znajek process, is negligible for solving the field equations. The physically more viable setting, force-free electrodynamics rather than electro-vacuum, would indeed probably increase the jet creation efficiency, see \cite{Komissarov-2007}. Hence, the idealized situation treated here, yields a lower bound.



The efficiency of the Blandford--Znajek process is given in geometrical units by \cite{Tchekhovskoy2010,Tchekhovskoy-2011}
\begin{align}\label{eq:efficiency formula}
\eta &= \frac{\varkappa}{4\pi}x^2 \left\langle\Phi^2_{\mathrm{BH}}\,( \dot{M}\,M^2)^{-1/2}\right\rangle\,(1+1.38\,x^2 - 9.2\,x^4),
\end{align}
where $\varkappa$ is a constant depending on the geometry of the magnetic field, $x$ is a variable given in terms of the dimensionless spin parameter $a/M$:
\begin{align}
x &= \frac{a/M}{2(1+\sqrt{1-(a/M)^2})}.
\end{align}
$ \Phi_{\mathrm{BH}}$ is the flux of the magnetic field through a hemisphere of the horizon, $\dot{M}$ is the accretion rate, and $\langle \dots \rangle$ is a time average.

\begin{figure*}
\begin{center}
\includegraphics[width=0.6\textwidth]{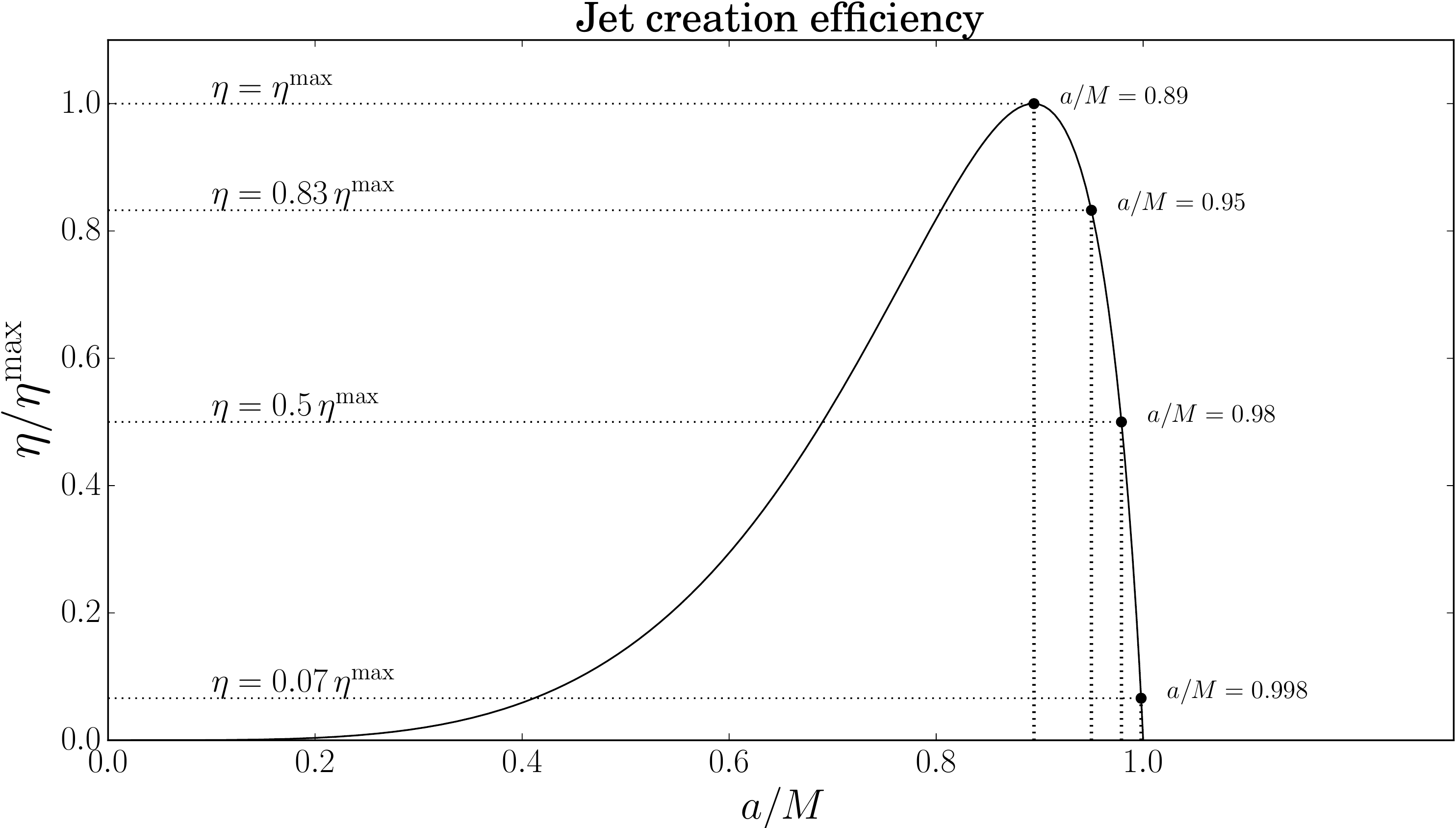}
\end{center}
\caption{{\bfseries Efficiency of the jet production}
${\eta}/{\eta}^{\max}$ depending on the
spin $a$, which varies from $0$ to $a=M$. The maximal value
${\eta}^{\max}$ is acquired for $a/M \approx 0.89$.}
\label{fig:efficiency}

\end{figure*}

In order to investigate the behavior of the jet production efficiency independently of a particular model of accretion, we vary the spin $a$ keeping all other parameters fixed. The result is depicted in Fig.\ \ref{fig:efficiency}, where we chose the same deformation as for Fig.\ \ref{fig:meissner}. Other deformations give qualitatively the same result. From Fig.\ \ref{fig:efficiency} any particular model can be recovered by a simple rescaling.

As Fig.\ \ref{fig:efficiency} shows, the efficiency is increasing up to $a/M
\approx 0.89$. For higher spins, the efficiency drops.
However, it deviates from the maximum value only by about 17\% for $a/M = 0.95$
and by 50\% for $a/M=0.98$. For even higher spins, it decays rapidly
to zero. Estimates of the maximal expected spin of a black hole with an accretion disk depend on the particular model chosen and ranges from
$a/M \approx 0.9$ to $a/M \approx  0.95$ for magnetohydrodynamic simulations of
thick disks \cite{Gammie2004,Benson2009}, which would still admit a high efficiency for the jet creation. For thin disks with low viscosity \cite{Sadowski2011} and for models taking only radiation into account \cite{Thorne1974} the limit can be as high as $a/M\approx 0.9994$ and $a/M \approx 0.998$, respectively.

Our result suggests that even if the Meissner effect would not be suppressed by the presence of matter crossing the horizon, it quenches the jet creation significantly only for black holes with spins higher than $a/M \approx 0.98$.

\begin{acknowledgments}
N.G.\ and M.S.\ thank the Friends of ZARM for their financial support and hospitality. M.S.\ was financially
supported by the Albert Einstein Center, grant no.\ 14-37086G by GA\v{C}R. N.G.\ gratefully acknowledges
support from the DFG within the Research Training Group 1620 ``Models of Gravity''. Partial support comes also from NewCompStar, COST Action MP1304. The authors
thank D.\ Giulini for helpful discussions. We are also grateful to the referees for valuable suggestions improving the paper.

We dedicate this work to Ji\v{r}\'i Bi\v{c}\'ak on the occasion of his 75th birthday. He pioneered the investigation of the Meissner effect during the
last decades and we are grateful to him for very enlightening discussions.

\end{acknowledgments}

\appendix

\section{Killing equations}\label{app:killing}

Let $K^a$ be a Killing vector of a spacetime. We expand it as
\begin{align}
 K_a &= K_0\,n_a + K_1\,\ell_a - K_2\,\bar{m}_a - \bar{K}_2\,m_a,\label{eq:Ka expansion}
\end{align}
so that the spin weights of $K_0, K_1, K_2$ and $\bar{K}_2$ are $0, 0, 1$ and $-1$, respectively. Then, the projections of the Killing equations
\begin{align}
 \nabla_aK_b + \nabla_b K_a = 0
\end{align}
onto the null tetrad read
\begin{subequations}
\begin{align}
D K_0 &= (\eps + \bar{\eps})K_0 - \bar{\kappa} K_2 - \kappa \bar{K}_2, \label{np:KE:DK0} \\
D K_1 + \Delta K_0  & = (\gamma+\bar{\gamma})K_0 - (\eps+\bar{\eps})K_1 + \nonumber\\
      &\quad + (\pi - \bar{\tau})K_2 + (\bar{\pi}-\tau)\bar{K}_2,\label{np:KE:DK1} \\
D K_2 + \delta K_0 &=  \bar{\pi}+\bar{\alpha}+\beta)K_0 - \kappa K_1 + \nonumber \\
 &\quad + (\eps-\bar{\eps}-\bar{\rho})K_2 - \sigma \,\bar{K}_2,\label{np:KE:DK2}\\
  \Delta K_1 &= - (\gamma+\bar{\gamma})K_1 + \nu \,K_2+\bar{\nu}\,\bar{K}_2,
               \label{np:KE:DeltaK1}
\end{align}
\begin{align}
  \Delta K_2 + \delta K_1 &=  \bar{\nu}\,K_0 - (\beta+\tau+\bar{\alpha})K_1 + \nonumber\\
      &  \quad + (\gamma-\bar{\gamma}+\mu)K_2 + \bar{\lambda}\bar{K}_2, \label{np:KE:DeltaK2}\\
 \eth K_2 &= \bar{\lambda} \,K_0 - \sigma\,K_1 ,
            \label{np:KE:deltaK2}\\
  \eth \bar{K}_2 + \bar{\eth}K_2 &= (\mu + \bar{\mu})K_0 - (\rho+\bar{\rho})K_1.\label{np:KE:deltaK2bar}
\end{align}
\label{np:KE}
\end{subequations}

In the paper we employ two Killing vectors. The stationary Killing vector $K^a$ reduces to $\ell^a$ on the horizon, i.e.\
\begin{align}
  K_0 &= 0, &
                   K_1 &= 1, &
                                    K_2 &= 0, \quad \text{on $\HH$},
\end{align}
and the Killing equation \eqref{np:KE:DeltaK1} together with Eq.\ \eqref{eq:gamma nu tau} implies $K_1 = 1$ everywhere in the neighborhood of the horizon.

The axial Killing vector $\eta^a$ satisfies the Killing equations \eqref{np:KE} in which $K$ has to be replaced by $\eta$ everywhere. On the horizon we
have $\eta^a = (\pd_\fii)^a$ and hence
\begin{align}
  \eta_0 &= \eta_1 = 0 \quad \text{on $\HH$}.
\end{align}
For the choice \eqref{eq:delta} we have
\begin{align}\label{eq:eta2}
  \eta_2 &= -\frac{\ii R\sin\theta}{\sqrt{2}} \quad \text{on $\HH$}.
\end{align}

\bibliography{references}

\end{document}